\documentclass{article}
\usepackage{
spconf,
amsmath,
graphicx,
cite,
multirow,
tabularx,
booktabs,
array,
amssymb,
enumitem,
amsmath,
url,
}

\title{Effective Noise-aware Data Simulation for Domain-adaptive Speech Enhancement Leveraging Dynamic Stochastic Perturbation}

\name{Chien-Chun Wang$^1$, Li-Wei Chen$^3$, Hung-Shin Lee$^3$, Berlin Chen$^1$, and Hsin-Min Wang$^2$}
\address{$^1$Dept. Computer Science and Information Engineering, National Taiwan Normal University, Taiwan\\
$^2$Institute of Computer Science, Academia Sinica, Taiwan\\
$^3$United-Link Co., Ltd., Taiwan}

\begin{document}
\maketitle

\begin{abstract}
Cross-domain speech enhancement (SE) is often faced with severe challenges due to the scarcity of noise and background information in an unseen target domain, leading to a mismatch between training and test conditions. This study puts forward a novel data simulation method to address this issue, leveraging noise-extractive techniques and generative adversarial networks (GANs) with only limited target noisy speech data. Notably, our method employs a noise encoder to extract noise embeddings from target-domain data. These embeddings aptly guide the generator to synthesize utterances acoustically fitted to the target domain while authentically preserving the phonetic content of the input clean speech. Furthermore, we introduce the notion of dynamic stochastic perturbation, which can inject controlled perturbations into the noise embeddings during inference, thereby enabling the model to generalize well to unseen noise conditions. Experiments on the VoiceBank-DEMAND benchmark dataset demonstrate that our domain-adaptive SE method outperforms an existing strong baseline based on data simulation.
\end{abstract}


\section{Introduction}\label{sec:intro}

Speech enhancement (SE) is crucial for improving speech signal quality and intelligibility by mitigating the detrimental impacts of background noise and interference. Recent years have witnessed remarkable advancements in SE driven by deep learning instantiated with models such as convolutional neural networks (CNNs) \cite{park2016,tan2019}, recurrent neural networks (RNNs) \cite{hu2018,zhao2018}, Transformer \cite{deOliveira2022}, generative adversarial networks (GANs) \cite{fu2019,xiang2020,cao2022,zadorozhnyy2022}, and many others. In particular, iconic supervised-learning approaches to SE typically rely on paired noisy and clean speech data to train their associated models that can transform a noisy speech signal into its clean (or enhanced) counterpart. However, these models often struggle with the domain mismatch problem caused by the inherent variability of real-world noise distributions that are difficult to fully render in training datasets. Consequently, the models trained on specific noise distributions may exhibit significant performance degradation when faced with unseen noise types or acoustic environments during deployment, highlighting a critical challenge in developing robust and widely-applicable speech enhancement systems.

To overcome the challenge of domain mismatch, researchers have explored various domain adaptation techniques \cite{liao2019,wang2015,long2016,tzeng2017,shu2019,hou2021}, aiming to bridge the gap between the noise distributions in the training and test conditions. These techniques, however, often involve complex training procedures or might fail to fully exploit the relationships between domains. Recently, data simulation has emerged as promising approaches for unsupervised noise adaptation in SE \cite{chen2023,chen2022a}. These approaches, unlike traditional adversarial training, directly synthesize noisy speech tailored to a target domain using clean speech. By leveraging the domain-invariant nature of clean signals, the associated models generate noisy speech that closely resembles those expected in the target domain, allowing for efficient model adaptation through fine-tuning on a large, simulated parallel dataset. Notably, these models exhibit high data efficiency, requiring only few minutes of unlabeled target-domain data to learn the noise distribution embodied in the spectrogram. Additionally, they operate without paired training data, focusing only on the target noise while deriving the clean-to-noisy transformation from non-matched utterances. However, existing data simulation approaches for SE mainly focus on replicating the overall spectral characteristics of the target-domain noise. They often fail to explicitly model and incorporate the intricate, fine-grained features of the noise, thereby potentially limiting the generalization ability of the trained SE models.

\begin{figure*}[ht]
\centering
\includegraphics[width=1.0\linewidth]{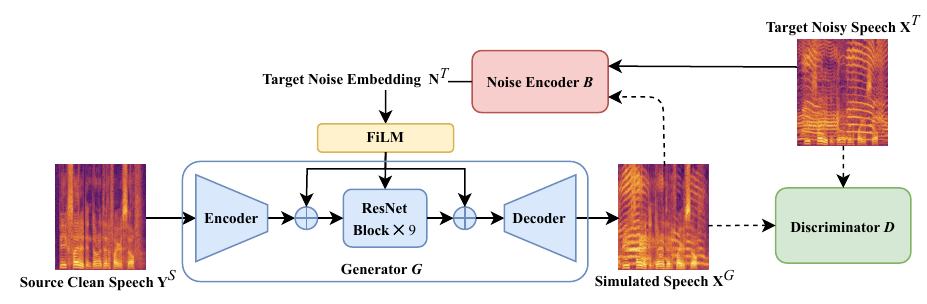}
\vspace{-25pt}
\caption{The architecture of our proposed method, NADA-GAN. The dotted arrows indicate that during the training phase, simulated speech $\mathbf{X}^G$ is used together with target noisy speech $\mathbf{X}^T$ to 1) train the discriminator, and 2) contribute to noise information reconstruction. The $\bigoplus$ operator denotes element-wise tensor addition.}
\label{fig:NADA-GAN}
\vspace{-10pt}
\end{figure*}

To address the limitation of existing data simulation approaches, we propose a novel Noise-Aware Domain-Adaptive method using Generative Adversarial Networks (dubbed NADA-GAN) for SE, specifically designed to capture and leverage the intricate noise characteristics of the target domain, even with limited data. Our method employs a dedicated noise encoder to extract rich noise embeddings directly from target-domain noisy speech. These embeddings then guide a generator network to synthesize noisy speech that faithfully replicates noise characteristics of the target domain meanwhile preserving the phonetic content of the input clean speech. Furthermore, we introduce the notion of dynamic stochastic perturbation during inference, which injects controlled perturbations into the extracted noise embeddings, facilitating the associated model to better generalize to a variety of unseen noise conditions beyond those observed during training. We fine-tune a \textit{causal} SE model with our generated data and evaluate our proposed method on the VoiceBank-DEMAND benchmark dataset, demonstrating its superiority over eminent approaches in terms of objective SE metrics. This study makes at least two significant contributions:

\begin{enumerate}[label=\arabic*),leftmargin=14pt,itemsep=0ex,topsep=4pt]
\item \textbf{Fine-grained Target Domain Simulation}: Unlike GAN-based approaches like UNA-GAN \cite{chen2023}, which ignore the differences in individual target utterances during simulation, our method enables the generation of simulated speech tailored to the specific enrolled target noisy speech. Furthermore, we introduce controlled variability in the noise embedding during simulation.
\item \textbf{Broad Applicability Beyond Speech Enhancement}: The superior performance of our method, as evidenced by mean opinion score (MOS) evaluation, demonstrates its versatile potential beyond SE. Moreover, it can be extended to train automatic speech recognition (ASR) models in a multi-conditional manner, suggesting possible avenues for further research in robust speech processing.
\end{enumerate}

\section{Proposed Method}\label{sec:method}

Figure \ref{fig:NADA-GAN} schematically depicts the architecture of our proposed method, NADA-GAN. The process starts with a generator $G$, which takes as input a clean magnitude spectrogram (hereafter referred to as spectrogram) $\mathbf{Y}^S$ from the source domain $S$ and generates a simulated spectrogram $\mathbf{X}^G$ that approximates the noise characteristics in the target domain $T$. To achieve this, the generator utilizes the target noise embedding $\mathbf{N}^T$ derived from the target noisy spectrogram $\mathbf{X}^T$ by the noise encoder $B$. The role of the discriminator $D$ is to distinguish the target noisy spectrogram $\mathbf{X}^T$ from the simulated spectrogram $\mathbf{X}^G$, thereby providing feedback to help enhance the generator's output. In addition, the noise reconstruction loss $\mathcal{L}_{nse}$ is employed to ensure accurate reconstruction of the noise embedding.

\subsection{Generator and Discriminator}\label{sec:G & D}

The generator $G$ transforms a clean spectrogram from the source domain into its simulated counterpart. Specifically, the clean spectrogram is first encoded through two 2D downsampling convolutional layers (kernel size: $3\times 3$, stride: $2\times 2$), followed by nine residual blocks for capturing deep representations. Each residual block consists of two convolutional layers (kernel size: $3\times 3$, stride: $1\times 1$) and a dropout layer. Finally, two transposed convolutional layers (kernel size: $3\times 3$, stride: $2\times 2$) act as a decoder, upsampling the representations into the corresponding simulated spectrogram.

The discriminator $D$ differentiates between the simulated spectrogram and the real target noisy spectrogram. It consists of five 2D convolutional layers (kernel size: $4\times 4$) followed by Leaky ReLU activation functions. The stride is set to $2\times 2$ for the first three layers and $1\times 1$ for the last two. We employ the adversarial loss during training:
\begin{equation}
\begin{split}
\mathcal{L}_{adv}(G,&D,\mathbf{X}^T,\mathbf{Y}^S,\mathbf{N}^T) = \mathbb{E}_{\mathbf{x} \sim \mathbf{X}^T} \left[ \log D(\mathbf{x}) \right] \\
&+ \mathbb{E}_{\mathbf{y} \sim \mathbf{Y}^S, \mathbf{n} \sim \mathbf{N}^T} \left[ \log (1-D(G(\mathbf{y},\mathbf{n}))) \right].
\end{split}
\end{equation}
This loss function encourages the generator to produce a spectrogram similar to the target noisy spectrogram. Although the speech content might differ, the discriminator here focuses exclusively on the background noise for discrimination.

\subsection{Noise Encoder}\label{sec:BEATs}

Drawing inspiration from recent advancements in noise-aware speech processing \cite{li2021,hu2024}, we propose the incorporation of a dedicated noise encoder, denoted by $B$ for our purpose. This encoder is tasked with extracting salient noise embeddings of the target noisy spectrogram, denoted by $\mathbf{N}^T$, from the outputs of its penultimate layer. By explicitly providing these noise representations to the generator module, our model gains a deeper understanding of the specific noise characteristics inherent to the target domain, enabling more informed and tailored noise synthesis.

To extract noise embeddings richly imbued with noise-specific information, we leverage a pre-trained audio model known as BEATs \cite{chen2022}. Pre-trained on the extensive AudioSet dataset \cite{gemmeke2017}, BEATs possesses a strong foundation in generating audio feature representations. To further enhance its noise discriminative capabilities, we fine-tune BEATs using a two-pronged approach: 1) We train BEATs to classify noise types on the VoiceBank-DEMAND dataset \cite{valentini-botinhao2016}. This dataset provides a diverse set of 10 distinct noise types, empowering the encoder to discern between a variety of noise characteristics. 2) We treat each utterance within the target noisy dataset used for GAN training as a unique noise type. This fine-tuning step further reinforces the noise encoder's ability to capture the specific noise nuances present in the data it will encounter during the speech enhancement process.

To ensure faithful noise information reconstruction, we introduce the noise reconstruction loss:
\begin{equation}\label{eq_nse}
\mathcal{L}_{nse}(G,\mathbf{Y}^S,\mathbf{N}^T) = \mathbb{E}_{\mathbf{y} \sim \mathbf{Y}^S, \mathbf{n} \sim \mathbf{N}^T} \left[ \left\| \mathbf{n} - B(G(\mathbf{y},\mathbf{n})) \right\|_1 \right].
\end{equation}
This loss function ensures that the generated samples retain the noise details present in the target domain.

\subsection{FiLM}\label{sec:film}

We adopt Feature-wise Linear Modulation (FiLM) \cite{perez2018} to incorporate noise embeddings into the generator module. At the outset, the noise embedding $\mathbf{N}^T$ undergoes separate linear transformations to produce weight $\mathbf{W}$ and bias $\mathbf{b}$:
\begin{equation}
\mathbf{W} = Linear(\mathbf{N}^T),\:\mathbf{b} = Linear(\mathbf{N}^T).
\end{equation}
Following this, the output feature $\mathbf{F}$ from a specific layer within the generator is multiplied by $\mathbf{W}$ and biased by $\mathbf{b}$: 
\begin{equation}
\mathbf{F}' = \mathbf{W} \times \mathbf{F} + \mathbf{b}.
\end{equation}
Finally, the adjusted feature $\mathbf{F}'$ propagates through the remaining layers of the generator. This approach is applied to the output features of the encoder layer and all nine ResNet blocks within the generator.

\subsection{Patch-wise Contrastive Learning}\label{sec:MPC}

To preserve linguistic consistency between the generated noisy speech and the original clean audio, we employ patch-wise contrastive learning, as introduced by Park \textit{et al.} \cite{park2020}, which homes in maximizing the mutual information, specifically the shared speech content information, between the source clean and the simulated spectrograms.

We utilize the generator to extract feature representations for both the clean and simulated spectrograms. Within these representation spaces, a small patch from the simulated representation is selected as the ``query''. The corresponding patch from the clean representation serves as the ``positive'' sample. To provide contrasting examples, we randomly sample 256 patches from the clean representation to act as ``negative'' samples. These patches are then projected into a lower-dimensional embedding space using two linear layers with 256 units each, followed by ReLU activation. The contrastive loss, calculated across five layers of the generator's feature hierarchy, measures the cross-entropy loss between the ``query'' patch and both the positive and negative patches. This formulation encourages the model to learn representations where corresponding patches in the clean and noisy domains exhibit high similarity, while simultaneously being easily distinguishable from randomly sampled patches. Formally, the patch-wise contrastive learning loss is defined by:
\vspace{-3pt}
\begin{equation}
\resizebox{0.5\textwidth}{!}{
$\mathcal{L}_{pcl}(G,\mathbf{Y}^S) = \sum\limits_{l=1}^{L} \sum\limits_{i=1}^{I} -\log \left[ \frac{e^{\left( \hat{z}_l^i \cdot z_l^i / \tau \right)}}{e^{\left( \hat{z}_l^i \cdot z_l^i / \tau \right)} + \sum_{j=1}^{J} e^{\left( \hat{z}_l^i \cdot z_l^j / \tau \right)}} \right]$,
}
\end{equation}
where $z_{l}^{i}$ represents the $i^{th}$ positive patch from clean representations at the $l^{th}$ layer of the generator, $\hat{z}_{l}^{i}$ denotes the corresponding patch from simulated representations, and $z_{l}^{j}$ refers to the $j^{th}$ negative patch from simulated representations at the same layer. The temperature parameter $\tau$ controls the contrastive learning process. We apply the loss function to both source clean ($\mathcal{L}_{pcl}(G,\mathbf{Y}^S)$) and target noisy ($\mathcal{L}_{pcl}(G,\mathbf{X}^T)$) spectrograms to allow for consistent speech content and minimize unnecessary alterations.

\subsection{Training Objective}\label{sec:train}
During the training phase, the main objective is to optimize the GAN model using a comprehensive loss function that integrates multiple components. The total loss function combines adversarial loss, patch-wise contrastive learning losses for both source clean and target noisy spectrograms, and the noise reconstruction loss, which is formulated by:
\begin{equation}\label{eq_total}
\begin{split}
\mathcal{L}_{total} = &\mathcal{L}_{adv}(G,D,\mathbf{X}^T,\mathbf{Y}^S,\mathbf{N}^T) + \mathcal{L}_{pcl}(G,\mathbf{Y}^S) \\
&+ \mathcal{L}_{pcl}(G,\mathbf{X}^T) + \lambda_{nse}\mathcal{L}_{nse}(G,\mathbf{Y}^S,\mathbf{N}^T),
\end{split}
\end{equation}
where $\lambda_{nse}$ is a hyperparameter that regulates the influence of the noise reconstruction loss.

Given limited noisy data from the target domain, the same amount of clean speech are randomly sampled from the source domain. Subsequently, our model is trained using the unpaired dataset and optimized with $\mathcal{L}_{total}$ in Eq. (\ref{eq_total}).

\subsection{Inference with Dynamic Stochastic Perturbation}\label{sec:infer}
After training, our well-trained generator serves as a domain converter $F^{S \sim T}$ from $\mathbf{Y}^S$ to $\mathbf{X}^T$, working alongside the fine-tuned noise encoder to simulate noisy speech. The simulation utilizes abundant clean speech data and randomly selected target noisy speech from the training phase, providing sufficient paired data of clean and simulated speech. This augmented data can be further utilized to fine-tune any downstream speech enhancement models for domain adaptation.

Dynamic stochastic perturbation is a technique used during the inference stage to improve the adaptability of speech enhancement models. It involves adding Gaussian noise, with adjustable standard deviation, to noise embeddings extracted from target noisy spectrograms. The standard deviation represents the different degrees of spread used for generating the stochastic perturbation. The dynamic noise injection allows for fine-tuning of noise levels, assisting in preventing overfitting to specific noise patterns encountered during training. Moreover, it enhances the resilience of speech enhancement models to unseen noise conditions, thereby bolstering their generalization capabilities.

\section{Experiments}\label{sec:exp}

\subsection{Dataset}\label{sec:dataset}
We evaluated our proposed NADA-GAN method on the VoiceBank-DEMAND dataset \cite{valentini-botinhao2016}, a widely used benchmark for speech enhancement. This dataset consists of noisy speech recordings generated by artificially mixing clean speech samples from the VoiceBank corpus \cite{valentini-botinhao2016} with noise samples from the DEMAND database \cite{thiemann2013}. It comprises recordings from 30 speakers, with 28 used for training and 2 reserved for testing. The training set (source domain) contains 11,572 utterances created by mixing clean speech with 10 noise types at four signal-to-noise ratio (SNR) levels (0, 5, 10, and 15 dB). Conversely, the test set (target domain) consists of 824 utterances featuring five unseen noise types at four SNR levels (2.5, 7.5, 12.5, and 17.5 dB).

To facilitate efficient training with limited target data, we randomly sampled 40 unpaired utterances: 40 clean speech utterances from the source domain and 40 noisy speech utterances from the target domain (8 utterances per noise type). Furthermore, we excluded the portion of the original test set used for training the data simulation models. Consequently, the final test set was reduced from 824 to 784 utterances.

\subsection{Speech Enhancement Backbone Model}\label{sec:DEMUCS}
We employed DEMUCS \cite{defossez2020} as our downstream evaluation model. DEMUCS is a real-time SE model that operates directly in the waveform domain. It utilizes a \textit{causal} encoder-decoder architecture with U-Net style skip connections and convolutional layers, enabling real-time processing. DEMUCS directly processes the raw waveform to suppress stationary and non-stationary noise, as well as room reverberations. The model optimizes for both time and frequency domain losses by incorporating an L1 loss on the waveform and a multi-resolution short-time Fourier transform (STFT) loss on the spectrogram. To further enhance robustness, data augmentation techniques, including frequency band masking and signal reverberation, are applied directly to the raw waveform during training. For comparison, we trained four variants of speech enhancement models:
\begin{enumerate}[leftmargin=0pt, label={}, itemsep=0pt,topsep=4pt]
\item \textbf{Vanilla-SE}: DEMUCS trained solely on source domain data without any domain adaptation.
\item \textbf{UNA-GAN} \cite{chen2023}: Vanilla-SE fine-tuned with a dataset generated by the baseline domain adaptation approach.
\item \textbf{NADA-GAN}: Vanilla-SE fine-tuned with a dataset generated by our proposed method.
\item \textbf{Oracle}: Vanilla-SE fine-tuned with a mixture of clean speech and the \textit{real} target noise used for GAN training.
\end{enumerate}

\subsection{Training and Inference for Simulation}

To better capture the short-time characteristics of noise, we segmented the magnitudes into frames with dimensions of $129 \times 128$. This segmentation allows the noise encoder to learn more detailed noise-specific information. We trained our method for 400 epochs to learn a robust mapping between source clean speech and target noisy speech. The hyperparameter $\lambda_{nse}$ in Eq. (\ref{eq_total}) was set to 10.0 based on empirical results, striking a balance between noise reconstruction accuracy and overall generative quality. We used the Adam optimizer \cite{kingma2017} with an initial learning rate of 0.0002 for training.

After training, we utilized the generator and the fine-tuned noise encoder to create a simulated dataset. Following the setup in \cite{chen2023}, we used 11,572 clean speech samples from the source domain and the same 40 target noisy samples used during training. For each clean sample, we generated a corresponding noisy sample using the generator. We applied dynamic stochastic perturbation with a given standard deviation during generation, resulting in a set of 11,572 paired clean and simulated utterances for SE model training.

\subsection{Adaptation of SE Models}

We fine-tuned the Vanilla-SE model on the simulated dataset for two epochs. This fine-tuning process adapted the model to the specific noise characteristics present in the simulated data, improving its ability to generalize to unseen noise conditions.

\begin{table}[t]
\small
\caption{PESQ and STOI results on VoiceBank-DEMAND.}
\vspace{5pt}
\label{tab:result_voicebank-demand}
\centering
\begin{tabular}{lcc}
\toprule
\bf{Model} & \bf{PESQ} & \bf{STOI (\%)} \\
\toprule
Noisy & 2.20 & 88.9 \\
Vanilla-SE & 3.05 & \textbf{95.2} \\
UNA-GAN \cite{chen2023} & 3.09 & 95.0 \\
\textbf{NADA-GAN} (Pre-trained BEATs) & 3.08 & 94.8 \\
\textbf{NADA-GAN} (Fine-tuned BEATs) & \textbf{3.14} & \textbf{95.2} \\
\midrule
Oracle & 3.11 & 95.1 \\
\bottomrule
\end{tabular}
\end{table}

\begin{figure}
\centering
\includegraphics[width=1.0\linewidth]{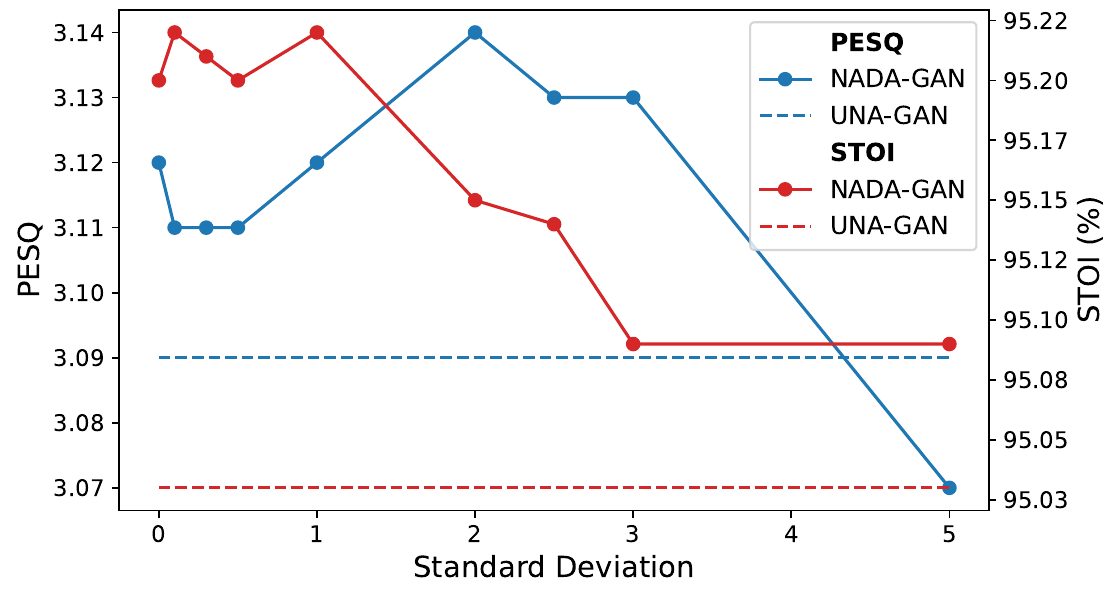}
\vspace{-15pt}
\caption{PESQ and STOI results of dynamic stochastic perturbation w.r.t. various standard deviations.}
\label{fig:result_dsp}
\vspace{-10pt}
\end{figure}

\section{Results}\label{sec:results}

\subsection{Main Results on VoiceBank-DEMAND}\label{sec:pesq}
Table \ref{tab:result_voicebank-demand} presents the main results on the VoiceBank-DEMAND dataset. Our proposed method outperforms the baseline approach in terms of perceptual evaluation of speech quality (PESQ) \cite{rix2001} and short-time objective intelligibility (STOI) \cite{taal2011}. These results highlight the effectiveness of integrating the noise encoder into the domain adaptation process.

Interestingly, utilizing pre-trained BEATs within our proposed method leads to decreased performance. This finding underscores the importance of fine-tuning the noise encoder for the specific target domain. Most notably, our method surpasses the performance achieved by training solely on real target noisy data. This result demonstrates the superior generalization capabilities of our method compared to simply adding real noise to clean speech during training.

\subsection{Effects of Dynamic Stochastic Perturbation}\label{sec:dsp}

Figure \ref{fig:result_dsp} illustrates the impact of varying the standard deviation of the dynamic stochastic perturbation on our model's performance. The results clearly demonstrate that perturbation, preventing model from overfitting with only 40 target noisy utterances, can improve performance, but the optimal standard deviation is crucial. Specifically, increasing the standard deviation initially enhanced PESQ scores, peaking at 3.14 with a standard deviation of 2.0. However, further increases beyond this point led to a decrease in PESQ scores. This suggests that excessive perturbation can degrade the perceived quality of the enhanced speech. Conversely, STOI scores remained relatively stable across different standard deviations, indicating that speech intelligibility is less sensitive to changes in perturbation intensity.

\subsection{Ablation Studies}\label{sec:ablation}

To assess the contribution of each component in our model, we conducted ablation studies, with results presented in Table \ref{tab:result_ablation}. Removing the noise reconstruction loss (- $\mathcal{L}_{nse}$) led to a slight decrease in performance. This suggests that while the noise reconstruction loss contributes to maintaining noise fidelity during the domain adaptation process, its overall impact on the final enhancement performance is relatively small.

However, eliminating the noise embeddings (- Embeddings) resulted in a more substantial performance drop. This finding underscores the crucial role of noise embeddings in capturing and transferring noise characteristics from the target domain, ultimately leading to improved generalization to unseen noise conditions.

\begin{table}[t]
\small
\caption{PESQ and STOI results of the ablation studies.}
\vspace{5pt}
\label{tab:result_ablation}
\centering
\setlength{\tabcolsep}{20pt}
\begin{tabular}{lcc}
\toprule
\bf{Model} & \bf{PESQ} & \bf{STOI (\%)} \\
\toprule
\textbf{NADA-GAN} & \textbf{3.14} & \textbf{95.2}\\
- $\mathcal{L}_{nse}$ & 3.11 & \textbf{95.2}\\
- Embeddings & 3.09 & \textbf{95.2}\\ 
\bottomrule
\end{tabular}
\end{table}

\begin{figure}
\centering
\includegraphics[width=1.0\linewidth]{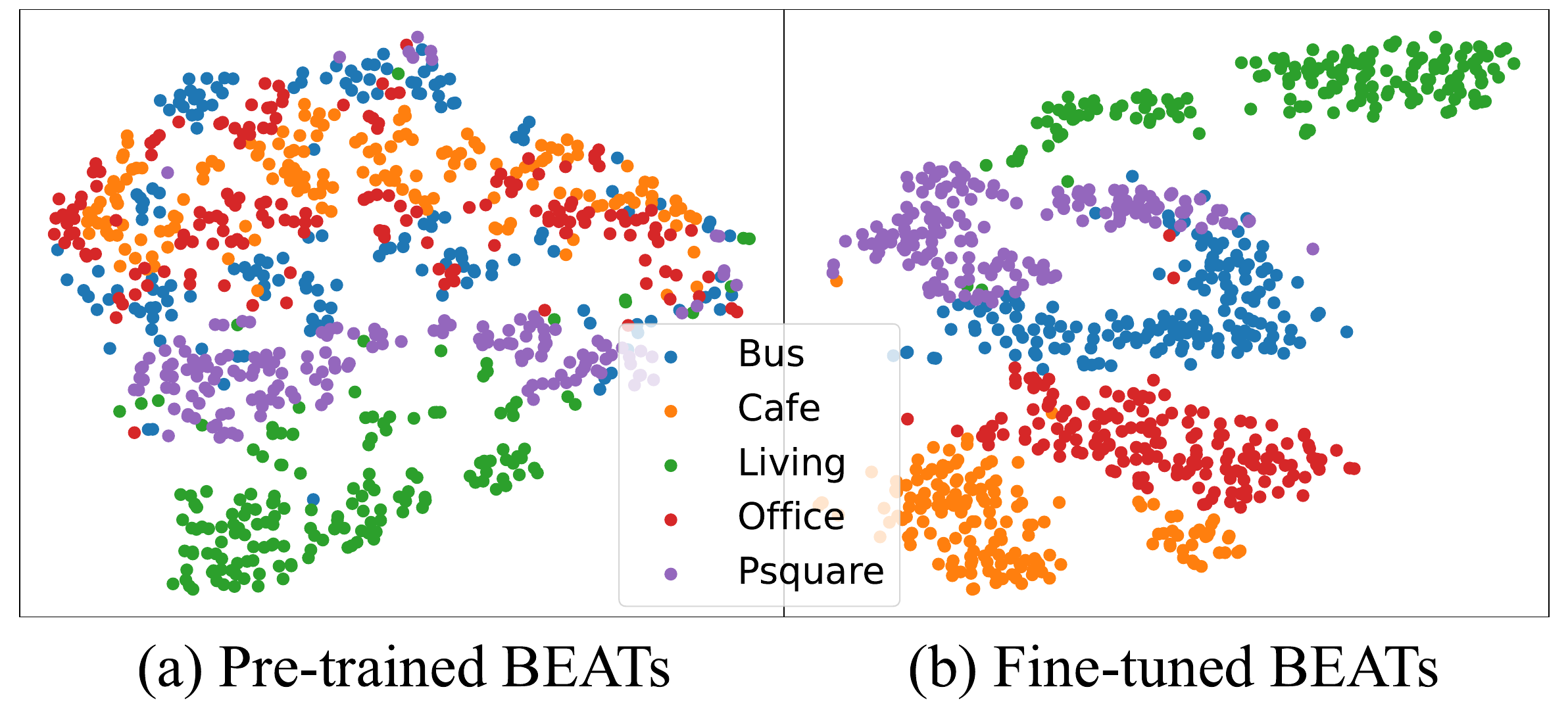}
\vspace{-15pt}
\caption{The t-SNE visualization of noise embeddings extracted from unseen non-stationary noise categories, i.e., ``Bus'', ``Cafe'', ``Living'', ``Office'', and ``Psquare''.}
\label{fig:t-SNE}
\vspace{-10pt}
\end{figure}

\subsection{t-SNE Visualization of Noise Embeddings}\label{sec:t-SNE}
To visualize the learned noise representations, we employed t-distributed stochastic neighbor embedding (t-SNE) \cite{maaten2008} on the noise embeddings extracted from the five unseen noise types in the VoiceBank-DEMAND test set. Figure \ref{fig:t-SNE} (a) and (b) display the embeddings obtained from the pre-trained BEATs and fine-tuned BEATs, respectively.

A clear separation between different noise categories is evident in the embeddings from the fine-tuned BEATs, even for non-stationary noise types. This improves discriminability, compared to the pre-trained embeddings, highlighting the effectiveness of fine-tuning in adapting the noise encoder to the target domain. The ability to distinguish between diverse noise profiles contributes significantly to the enhanced performance of our proposed method.

\subsection{MOS Evaluation on Simulated Data}\label{sec:mos}
To further evaluate the perceptual similarity of the generated noisy speech to the target domain, we conducted a MOS evaluation focusing on the non-stationary noise types. Table \ref{tab:result_MOS} presents the results. Ten participants, representative of typical listeners, participated in the evaluation. Each participant listened to 10 audio samples per model through headphones in a quiet room, using the target noisy speech as a reference. They were asked to rate the similarity of the background noise in each generated sample to the reference on a scale from 1 to 5, with 5 indicating the highest similarity. The MOS for each model was then calculated by averaging the scores across all participants and all samples.

The results demonstrate that our proposed model achieves a significantly higher MOS compared to the baseline model. This indicates that our method generates noisy speech with background characteristics that more closely resemble the target domain data, as perceived by human listeners.

\begin{table}[t]
\small
\caption{MOS evaluation of noisy speech generated by our method and the baseline approach.}
\vspace{5pt}
\label{tab:result_MOS}
\centering
\setlength{\tabcolsep}{32pt}
\begin{tabular}{lc}
\toprule
\bf{Model} & \bf{MOS} \\
\toprule
UNA-GAN \cite{chen2023} & 1.49 $\pm$ 0.42 \\
\textbf{NADA-GAN} & \textbf{2.51 $\pm$ 0.62} \\
\bottomrule
\end{tabular}
\vspace{-10pt}
\end{table}

\subsection{Results on Various SNR Levels}\label{sec:snr}
To evaluate the robustness of our method under different noise conditions, we assessed the performance of all models across various SNR levels. Table \ref{tab:result_snr} presents the results.

As expected, the baseline model outperforms Vanilla-SE, particularly at lower SNR levels, highlighting the effectiveness of its noise reduction capabilities. However, our proposed method consistently achieves the highest performance across all SNR levels, demonstrating significant and consistent improvements over the baseline. These results demonstrate that our proposed method is the most effective for speech enhancement under a range of noise conditions.

\subsection{SNR Distribution Analysis and Insights}\label{sec:snr_dis}
\begin{figure}
\centering
\includegraphics[width=1.0\linewidth]{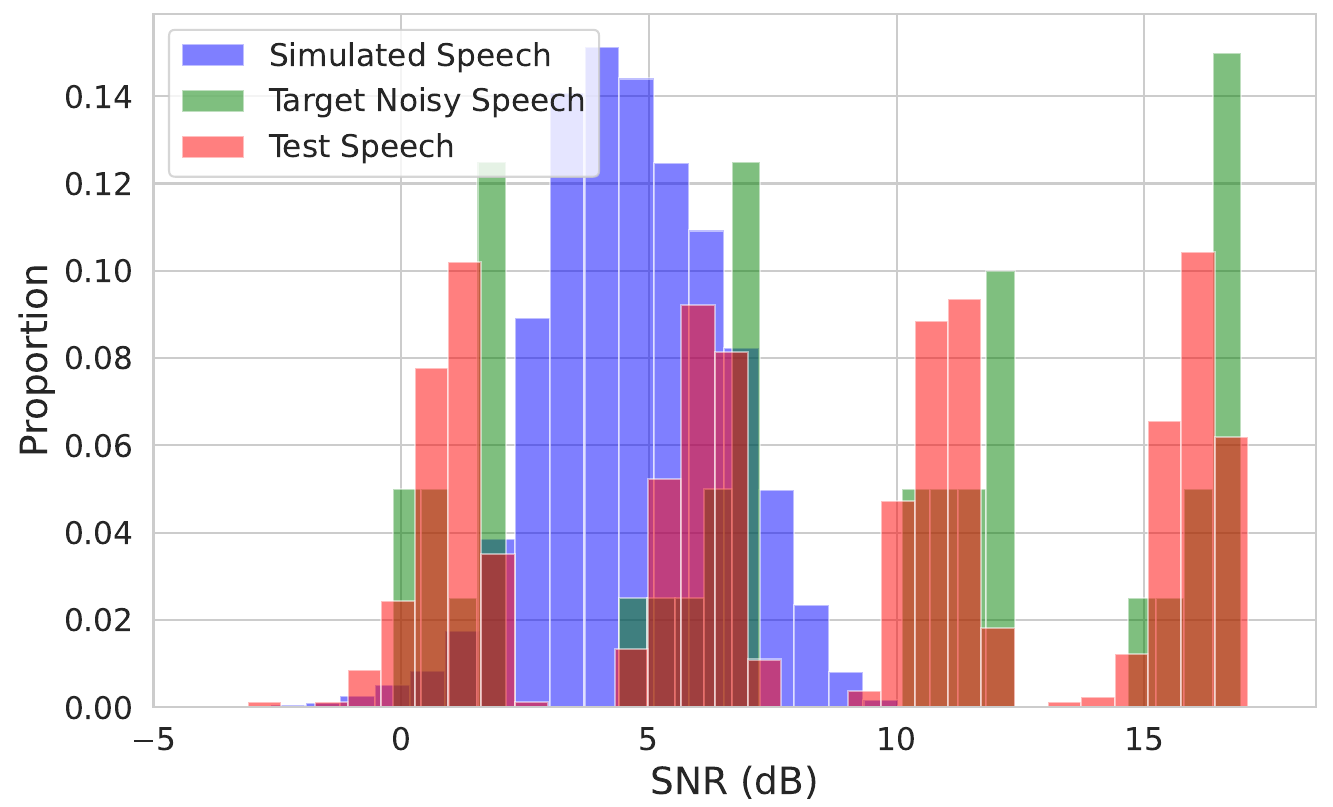}
\vspace{-20pt}
\caption{SNR distributions w.r.t. the simulated speech (11,572 utt.), target noisy speech (40 utt.) used for GAN training, and test speech (824 utt.) of VoiceBank-DEMAND.}
\label{fig:snr_dis}
\vspace{-10pt}
\end{figure}

Figure \ref{fig:snr_dis} presents the SNR distributions of the VoiceBank-DEMAND dataset. Notably, the SNR distribution of the target noisy speech closely resembles that of the test speech, with peaks around 0 dB and 15 dB SNR levels. This alignment between the target noisy speech and test speech corresponds to our data manipulation described in Section \ref{sec:dataset}.

In contrast, the simulated speech for the SE models exhibits a peak predominantly between 0 dB to 5 dB SNR. This concentrated distribution does not adequately reflect the broader SNR spectrum present in the test speech or real-world conditions. This observation highlights an area for improvement in future work: developing methods to generate simulated speech that encompasses and simulates a wider range of real-world SNR conditions would likely lead to further performance gains.

\begin{table}[t]
\small
\caption{PESQ results on VoiceBank-DEMAND w.r.t. various SNR levels.}
\vspace{5pt}
\label{tab:result_snr}
\centering
\setlength{\tabcolsep}{10.5pt}
\begin{tabular}{lcccc}
\toprule
\multirow{2}{*}{\textbf{Model}} & \multicolumn{4}{c}{\textbf{SNR (dB)}} \\ \cmidrule(lr){2-5}
 & \textbf{2.5} & \textbf{7.5} & \textbf{12.5} & \textbf{17.5} \\ 
\toprule
Vanilla-SE & 2.48 & 2.98 & 3.19 & 3.45 \\ 
UNA-GAN \cite{chen2023} & 2.54 & 3.00 & 3.22 & 3.48 \\ 
\textbf{NADA-GAN} & \textbf{2.59} & \textbf{3.05} & \textbf{3.27} & \textbf{3.53} \\ 
\bottomrule
\end{tabular}
\end{table}

\section{Conclusion and Future Work}
\label{sec:conclu}

This study\footnote{Code: \url{https://github.com/JethroWangSir/NADA-GAN/}.} introduces a novel approach to cross-domain speech enhancement that leverages noise-extractive techniques and generative adversarial networks (GANs). By incorporating a noise encoder and dynamic stochastic perturbation, our method effectively addresses the challenge of domain mismatch. This leads to improved generalization capabilities, enabling speech enhancement models to perform effectively across diverse and unseen noise conditions.

Experimental results on the VoiceBank-DEMAND dataset demonstrate the effectiveness of our proposed method. We achieve superior performance compared to a strong existing baseline across both objective and subjective metrics, including PESQ, STOI, and MOS. These findings underscore the effectiveness of our approach in enhancing both the quality and robustness of speech enhancement in real-world scenarios. In our future work, we plan to validate our methods on more advanced SE models across various challenging environments and datasets to thoroughly assess its effectiveness.

\bibliographystyle{IEEEbib}
\bibliography{references}

\end{document}